\documentclass[aps,pre,superscriptaddress,amsmath,preprint,showpacs]{revtex4}
\usepackage{amsmath}
\usepackage[dvips]{graphicx}
\begin{document}
\title{Numerical study of the magnetic electron drift vortex mode turbulence in a nonuniform magnetoplasma}

\author{Dastgeer Shaikh}
\affiliation{
Department of Physics and Center for Space Plasma and Aeronomic Research (CSPAR)\\
The University of Alabama in Huntsville,
Huntsville. Alabama, 35899}

\author{B. Eliasson}
\affiliation{Institut f\"ur Theoretische Physik IV, Ruhr-Universit\"at Bochum\\ D-44780 Bochum, Germany}
\affiliation{Department of Physics, Ume{\aa} University, SE-901 87, Ume{\aa}, Sweden}

\author{P. K. Shukla}
\affiliation{Institut f\"ur Theoretische Physik IV, Ruhr-Universit\"at Bochum\\ D-44780 Bochum, Germany}
\affiliation{Department of Physics, Ume{\aa} University, SE-901 87, Ume{\aa}, Sweden}
\affiliation{Scottish Universities Physics Alliance, Department of Physics,
University of Strathclyde, Glasgow G4 0NG, Scotland, U. K.}
\affiliation{GoLP/Instituto de Plasmas e Fusao Nuclear,
Instituto Superior T\'ecnico, 1049-001 Lisboa, Portugal}
\affiliation{School of Physics, University of KwaZulu-Natal, Durban 4000, South Africa}


\begin{abstract}
A simulation study of the magnetic electron drift vortex (MEDV) mode
turbulence in a magnetoplasma in the presence of inhomogeneities in
the plasma temperature and density, as well as in the external
magnetic field, is presented. The study shows that the influence of
the magnetic field inhomogeneity is to suppress streamer-like
structures observed in previous simulation studies without background
magnetic fields.  The MEDV mode turbulence exhibits non-universal
(non-Kolmogorov type) spectra for different sets of the plasma
parameters. In the presence of an inhomogeneous magnetic field, the
spectrum changes to a 7/3 power law, which is flatter than without
magnetic field gradients.  The relevance of this work to laser-produced 
plasmas in the laboratory is briefly mentioned.
\end{abstract}
\received{19 January 2008}
\maketitle
\section{Introduction}

The generation of magnetic field in laser-produced plasmas in the
laboratory
\cite{Stamper71,Raven78,Stamper91,Tatarakis02a,Tatarakis02b,Wagner04}
and in the universe
\cite{Grasso01,Kronberg03,Schlickeiser03,Kulsrud08,Ryu08} is an
fascinating and rich field of research.  Possible mechanisms to
spontaneously generate magnetic fields in a plasma include the
Biermann battery \cite{Biermann50}, which is associated with
non-parallel density and temperature gradients, and the Weibel
instability \cite{Weibel59}, in which the electrons have a
non-isotropic temperature.  The Weibel instability may be responsible
for the generation of large-scale magnetic fields in the Universe, as
well as in inertial fusion plasmas
\cite{Silva02,Fonseca03,Medvedov06}, while the Biermann battery has
been proposed as a possible mechanism to produce mega-Gauss magnetic
fields in laser-produced plasmas \cite{Stamper71}.  There are recent
observations of megagauss-field topology changes and structure
formation in laser-produced plasmas \cite{seguin}. The observations of
spontaneous generation of magnetic fields in plasmas
\cite{Stamper71,Raven78} and effects attributed to self-generated
magnetic fields, such as transport of energy along surfaces
\cite{Forslund82} and the insulation of laser-heated electrons from
the target interior \cite{Forslund82,Max78}, inspired the
investigation of magnetic surface waves in plasmas
\cite{Jones83}. This model was generalized to investigate the
generation of magnetic fields from temperature and density gradients
in the plasma \cite{Yu85}. The nonlinear properties of MEDV modes in a
nonuniform plasma with the equilibrium density and electron
temperature gradients have been investigated analytically without
\cite{Nycander87,Nycander91}, and with \cite{Murtaza89} background
magnetic fields. The spectral and statistical properties
\cite{Pavlenko94}, as well as the generation of large scale magnetic
fields \cite{Andrushchenko04} by the MEDV mode turbulence, have been
investigate both theoretically and numerically. The generation of
steep, non-Kolmogorov spectra of streamer-like structures by MEDV
turbulence has been observed in recent simulation studies
\cite{Shaikh08}.

In this paper, we report a simulation study of the MEDV mode
turbulence in the presence of gradients in the equilibrium electron
temperature and electron density, as well as in the background
magnetic field.  We concentrate our study on the magnetic field effect
and its influence on the spectral properties of the MEDV mode
turbulence. The MEDV mode turbulence involve a competition between
nonlinear zonal flows \cite{Shukla81}, which we define as nonlinear
MEDV modes with a finite scale in the direction of the equilibrium
plasma density and temperature gradients, and streamers that have a
finite scale perpendicular to the plasma gradients. We will here
assume that the background magnetic field gradient is in the same
direction as the plasma gradients. Our work presented here ignores
collisional effects. The latter are discussed in Ref. \cite{Yu89} that
includes frictional and thermal forces, as well as gradients of the
unperturbed magnetic field.

The manuscript is organized in the following fashion. The governing
nonlinear equations for the two-dimensional (2D) MEDV modes are
presented in Sec. II. The results of computer simulations are
displayed in Sec. III., and the observed spectral properties of the
MEDV turbulence are discussed in Sec. IV.  Finally, the results are
briefly summarized and discussed in Sec V.

\section{Nonlinear equations}

We here derive the governing equations for nonlinearly interacting 2D MEDV modes
in an inhomogeneous plasma containing equilibrium electron density, electron temperature
and background magnetic field gradients.  Following Ref.~\cite{Nycander87}, we will assume a 2D
geometry in the $xy$ plane, where the perturbed magnetic field is directed
along the $z$ axis so that ${\bf B}=B(x,y,t) \widehat{\bf z}$, where
$\widehat{\bf z}$ is the unit vector along the $z$ axis.
The governing equations for the wave magnetic field $B$
and electron temperature fluctuations $T_1$ are given by \cite{Nycander87}
\begin{equation}
  \frac{\partial}{\partial t}\left(
  \frac{e}{m}B-\frac{1}{\mu_0 e n_0}\nabla^2 B\right)
  =\frac{1}{(\mu_0 e n_0)^2}\{B,\nabla^2 B\} -\frac{1}{m n_0}\{n_0,T_1\}
  \label{EMHD}
\end{equation}
and
\begin{equation}
  \frac{\partial T_1}{\partial t}+\frac{1}{\mu_0 e n_0}\{B,T_0+T_1\}
  +(\gamma-1)\frac{T_0}{\mu_0 e n_0^2}\{n_0,B\}=0,
  \label{Energy}
\end{equation}
respectively. Here $e$ is the magnitude of the electron charge, $m$
is the electron mass, $\mu_0=1/c^2\epsilon_0$ is the magnetic permeability in vacuum,
$c$ is the speed of light, $\epsilon_0$ the electric permittivity
in vacuum, and $\gamma=5/3$ is the ratio between the specific heats. The equilibrium electron
number density $n_0$ and the electron temperature $T_0$ are assumed to have a gradient along the $x$ axis.
In Eqs. (\ref{EMHD}) and (\ref{Energy}), we have introduced the Poisson bracket
notation
\begin{equation}
  \{f,g\}=\frac{\partial f}{\partial x}\frac{\partial g}{\partial y}
  -\frac{\partial f}{\partial y}\frac{\partial g}{\partial x}.
\end{equation}
We have neglected the scalar nonlinearity \cite{Nycander87} in Eq. (\ref{EMHD}), which becomes important only on timescales much larger than the electron gyro period \cite{Nycander87}. On a longer timescale the effect of the scalar nonlinearity can be important, and numerical simulations \cite{Mikhailovskaya86} indicate that it causes dipolar vortices to gradually transform to monopolar vortices.

Dividing the magnetic field as $B=B_0+B_1$, where $B_0$ is the large-scale
background magnetic field and $B_1$ is the perturbations, and assuming
that the equilibrium quantities $n_0$, $T_0$ and $B_0$ depend only on
the coordinate $x$, we have
\begin{equation}
  \frac{\partial}{\partial t}\left(
  \frac{e}{m}B_1-\frac{1}{\mu_0 e n_0}\nabla^2 B_1\right)
  =\frac{1}{(\mu_0 e n_0)^2}(\{B_0,\nabla^2 B_1\}+\{B_1,\nabla^2 B_1\})
  -\frac{1}{m n_0}\{n_0,T_1\}
  \label{EMHD2}
\end{equation}
and
\begin{equation}
  \frac{\partial T_1}{\partial t}+\frac{1}{\mu_0 e n_0}(\{B_0,T_1\}+\{B_1,T_1\})
  +(\gamma-1)\frac{T_0}{\mu_0 e n_0^2}\{n_0,B_1\}=0,
  \label{Energy2}
\end{equation}
where we have used that $\{B_0,T_0\}=\{n_0,B_0\}=\{B_0,\nabla^2 B_0\}=0$.

In order to investigate the basic properties of the nonlinear system
of equations (\ref{EMHD2}) and (\ref{Energy2}), it is convenient to scale
it with its typical length and time scales. Noting that the characteristic length scale
of the system is the electron skin depth $\lambda_e = c/\omega_{pe}$,
where $\omega_{pe}=(n_0 e^2/\epsilon_0 m)^{1/2}$ is the electron plasma frequency,
and that the typical time scale is the plasma
inhomogeneity length scale divided by the thermal speed of the electrons \cite{Nycander87},
it is possible to cast the system (\ref{EMHD2})--(\ref{Energy2}) into the dimensionless form
\begin{equation}
  \frac{\partial}{\partial t}(B_1-\nabla^2 B_1)=\{B_1,\nabla^2B_1\}
  +\kappa_B\nabla^2\frac{\partial B_1}{\partial y}
  -\frac{\partial T_1}{\partial y},
  \label{B_norm}
\end{equation}
and
\begin{equation}
  \frac{\partial T_1}{\partial t}=-\{B,T_1\}-\kappa_B\frac{\partial T_1}{\partial y}
  -\sigma\frac{\partial B_1}{\partial y}.
  \label{T1_norm}
\end{equation}
where we have normalized the spatial coordinates $x$ and
$y$ by the electron skin depth $\lambda_e$, the time by
$\lambda_e/V_{Te}(|\kappa|\kappa_n)^{1/2}$, the magnetic field
by $(m/e) V_{Te}(|\kappa|\kappa_n)^{1/2}/\lambda_e$,
and the temperature fluctuations $T_1$ by $(|\kappa|\kappa_n)^{1/2}T_0$,
where $V_{Te}=(T_e/m)^{1/2}$ is the electron thermal speed.
The normalized background plasma gradients are given by $\kappa_n=(n_0'/n_0)(c/\omega_{pe})$,
$\kappa=[(\gamma-1)n_0'/n_0-(T_0'/T_0)](c/\omega_{pe})$, and the normalized magnetic field gradient by
$\kappa_B=(B_0'/B_0)(c/\omega_{pe})$, where the primes denote differentiation with respect to $x$.
The coordinate system is chosen such that $\kappa_n>0$.
With this normalization, $\sigma=+1$ for $\kappa>0$ and $\sigma=-1$ for $\kappa<0$.
Hence, the only parameters in (\ref{B_norm}) and (\ref{T1_norm}) are $\sigma$ and $\kappa_B$, where
$\sigma$ only takes the values $+1$ or $-1$.

Linearizing the system of Eqs. (\ref{B_norm}) and (\ref{T1_norm}),
and assuming that $B_1$ and $T_1$ are proportional to $\exp(ik_x x+ik_y y-i\omega t)$,
we obtain the linear dispersion relation
\begin{equation}
  (\omega-\kappa_B k_y)[\omega+k^2(\omega-\kappa_B k_y)]=k_y^2\sigma.
  \label{disp}
\end{equation}
where $\omega$ and ${\bf k} =\widehat{\bf x} k_x + \widehat{\bf y} k_y$ are the frequency and
wave vector, respectively. Equation (\ref{disp}) has solutions of the form
\begin{equation}
  \omega=\frac{k_y}{2(1+k^2)}\left[\kappa_B(1+2k^2)\pm\sqrt{\kappa_B^2+4(1+k^2)\sigma}\right].
\end{equation}
For $\sigma=-1$, we have unstable MEDV modes for $4(1+k^2)>\kappa_B^2$ when $k_y\neq 0$. For $\sigma=+1$, we have only stable MEDV modes, and one can find zero-frequency waves when $k=1/\kappa_B$, in addition to the zero frequency zonal flows at $k_y=0$.
In the limit $\kappa_B\rightarrow 0$ we retain the previous result \cite{Yu85,Nycander87}
\begin{equation}
  \omega=\pm \frac{k_y}{\sqrt{1+k^2}}\sqrt{\sigma},
\end{equation}
which predicts stable MEDV modes for $\sigma>0$ and purely growing MEDV modes for $\sigma<0$.
The unstable case $\sigma<0$ corresponds to a situation where density and temperature gradients are in the same direction and $(T_0'/T_0)>(\gamma-1)n_0'/n_0$. This instability gives rise to the generation of magnetic field fluctuations and is related to the first order baroclinic ($\nabla n_0\times \nabla T_1$) effect, which shows the importance of the temperature fluctuations for the instability to take place \cite{Yu85}.

The nonlinear system possesses the conserved energy integral
\begin{equation}
  {\cal E}=\int\int\left[B^2+(\nabla B)^2+\frac{T_1^2}{\sigma}\right]\,dx\,dy.
  \label{integral}
\end{equation}
We note that the total energy is independent of $\kappa_B$. For $\sigma=+1$ the energy integral is positive definite,
and thus does not allow the growing of large amplitude magnetic fluctuations from small-amplitude noise.
For $\sigma=-1$, however, the energy integral is non-definite and allows large-amplitude waves to grow from
the linear instability discussed above.

\section{Numerical study}

We have adapted the nonlinear fluid code that was developed to study
the evolution of MEDV modes in \cite{Shaikh08} in the presence of equilibrium
electron density and temperature gradients, to also
include a background gradient in the magnetic field.
Consequently, the energy stored in these gradients
excite linear as well as nonlinear instabilities in a different
manner, than that described in \cite{Shaikh08}, because the nonlinear
mode coupling interactions are significantly modified by the
presence of both plasma and magnetic field gradients.  Our
numerical code employs a doubly periodic spectral discretization of
magnetic field and temperature fluctuations in terms of its Fourier
components, while nonlinear interactions are de-convoluted back and
forth in real and Fourier spectral spaces. The time integration is
performed by using the 4th-order Runge-Kutta method.  A fixed time
integration step is used.  The conservation of energy given by (\ref{integral}) is used
to check the numerical accuracy and validity of our numerical code
during the nonlinear evolution of the magnetic field and temperature
fluctuations.  Varying spatial resolution (from $128^2$ to $512^2$),
time step ($10^{-2}, 5\times 10^{-3}, 10^{-3}$), constant values of
$K_n \lambda V_T^ 2/c^ 2 =0.1$ and $(2/3)K_n \lambda - K_T \lambda
=0.1$ are used to ensure the accuracy and consistency of our nonlinear
simulation results.  We also make sure that the initial fluctuations
are isotropic and do not influence any anisotropy during the
evolution. Nonlinear interactions can, however, lead to anisotropic
turbulent cascades by migrating spectral energy in either $k_y \approx
0$ or $k_x \approx 0$ modes. The preference of energy transfer in
either of the modes is determined primarily by the background
gradients in magnetic or temperature and density fields that, in nonlinearly
saturated state, govern the mode coupling interactions.  In our
present simulations, we largely focus on two cases, viz (i) small
amplitude and (ii) large amplitude evolution of turbulent
fluctuations. These two cases are characteristically different from
each other in terms of dominant nonlinear interactions.
To compare the effects of the amplitude we need to compare the
nonlinear terms (the vector nonlinearities) with the linear terms, most importantly the
coupling terms between the magnetic field and temperature fluctuations. Hence, we can compare
the nonlinear term $\{B_1,\nabla^2B_1\}$ to the coupling term ${\partial T_1}/{\partial y}$ in the dimensionless Eq. (6). In the simulations (Figs. 1 and 2) we observed that the amplitudes of the dimensionless magnetic field and temperature fluctuations are approximately the same for both the small and large amplitude cases when the turbulence is fully developed. Hence we may use $T_1\sim B_1$ and compare the nonlinear term  $\{B_1,\nabla^2B_1\}$ to ${\partial B_1}/{\partial y}$. As an order of magnitude estimate, the spatial derivatives are of the order $1/L$, where $L$ is the typical length scale of the fluctuations. Hence we have the order of magnitude estimates $|\{B_1,\nabla^2B_1\}|\sim (1/L)^4 |B_1|^2$ and $|{\partial B_1}/{\partial y}|\sim (1/L) |B_1|$. One can conclude that if $B_1\ll L^3$ then the linear coupling term dominates and we have a case of weak turbulence, while if $B_1\approx L^3$ or larger, then we have a fully nonlinear case with strong turbulence. In our simulations, the fluctuations have typical length-scales of $L\approx 1$  (corresponding to the electron skin depth in dimensional units) so that $|B_1|\ll 1$ constitutes a weakly nonlinear case while $B\sim 1$ is a strongly nonlinear case. In dimensional units, $B_1$ should be replaced by $(eB_1/m)\lambda_e/[V_{Te}(|\kappa|\kappa_n)^{1/2}]$ and $L$ by $L/\lambda_e$ so that $e|B_1|/m$ should be compared to $(V_{Te}/\lambda_e)(|\kappa|\kappa_n)^{1/2}(L/\lambda_e)^3$ in dimensional units. Since $L/\lambda_e\approx 1$, we then have the weakly nonlinear case $e|B_1|/m\ll (V_{Te}/\lambda_e)(|\kappa|\kappa_n)^{1/2}$ and the strongly nonlinear case $e|B_1|/m\approx (V_{Te}/\lambda_e)(|\kappa|\kappa_n)^{1/2}$ in dimensional units.

\begin{figure}
\begin{center}
\includegraphics[width=15.cm,height=12.cm]{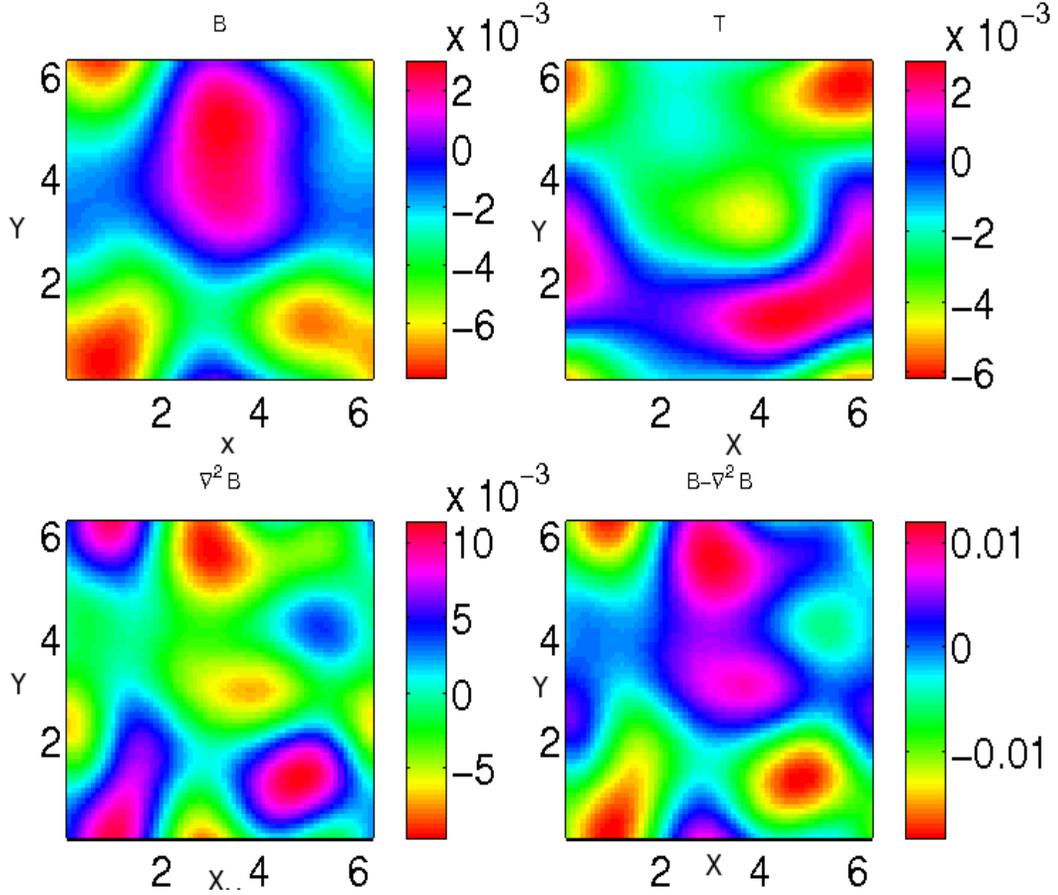}
\end{center}
\caption{\label{fig1} (Color online) Evolution of random turbulent fluctuations
  initialized with small amplitude. Nonlinear turbulent interactions
  lead to the formation of relatively large-scale flow directed along
  the mean magnetic field in our 2D simulations. The saturated
  structures in $B$, $T$, $\nabla^2 B$ and $B-\nabla^2 B$ are shown in
  the figure. The numerical resolution is $256^2$, box
dimension is $2\pi \times 2\pi$, and the parameters used are $\sigma =¨+1$ and $\kappa_B=0.5$.}
\end{figure}

\begin{figure}[t]
\begin{center}
\includegraphics[width=15.cm,height=12.cm]{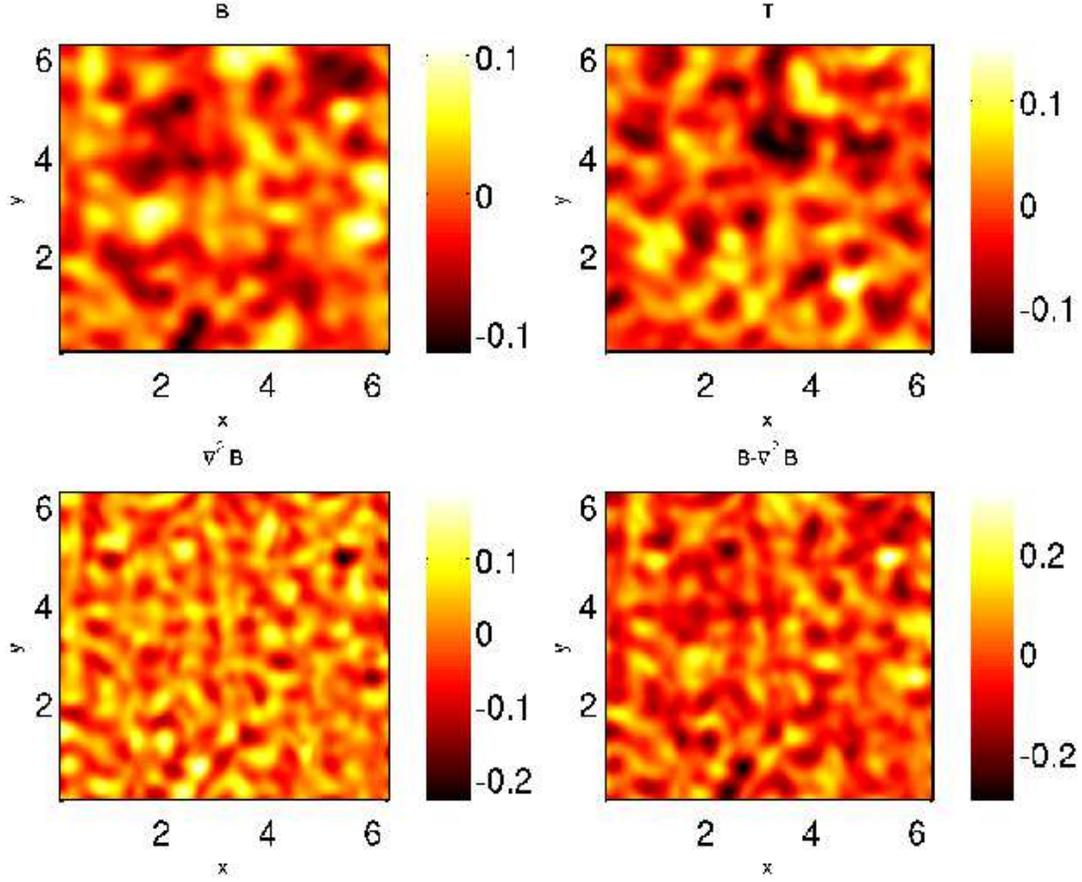}
\end{center}
\caption{\label{fig2} (Color online) When the amplitude of initial fluctuations in
  magnetic and temperature fields is large enough, characteristic
  nonlinear interaction modify the flows observed in Fig (1). The
  diamagnetic-like nonlinear interactions seem to dominantly suppress
  the flow and lead to small scale isotropic turbulent
  fluctuations. This is shown in $B$, $T$, $\nabla^2 B$ and
  $B-\nabla^2 B$. }
\end{figure}

The initial spectral distribution in the magnetic and temperature
fluctuations comprises a uniform isotropic and random amplitude
associated with the Fourier modes confined to a smaller band of wave
number ($k<0.1~k_{max}$). While spectral amplitude of the fluctuations
is random for each Fourier coefficient; it follows a $k^{-1}$ or
$k^{-2}$ scaling.  Note again that our final results do not depend on
the choice of the initial spectral distribution. The spectral
distribution set up in this manner initializes random scale turbulent
fluctuations. Since there is no external driving mechanism considered
in our simulations, turbulence evolves freely under the influence of
self-consistent nonlinear interactions. Note however that driven turbulence in
the context of the MEDV mode will not change inertial range spectrum
to be described here. The driving mechanism helps sustain turbulent
interactions without modifying the inertial range turbulent cascades.
The initial isotropic fluctuations in magnetic and temperature fields
are evolved through nonlinear fluid Eqs. (\ref{B_norm}) \& (\ref{T1_norm}).
The dominant nonlinear interactions in the inhomogeneous MEDV modes are governed by
$\hat{z}\times \nabla B \cdot \nabla \nabla^2 B$ in magnetic field
equation. This nonlinearity is similar to the polarization drift
nonlinearity $\hat{z}\times \nabla \phi \cdot \nabla \nabla^2 \phi$,
$\phi$ being the electrostatic potential fluctuations, in a two
dimensional Hasegawa-Mima-Wakatani (HMW) model describing drift waves
in inhomogeneous plasmas \cite{r13,r14,r15,r16}. This nonlinearity
characterizes Reynolds stress forces that plays a critical role in the
formation of zonal-flows. Analogously, one can expect generation of
nonlinearly generated flows in underlying MEDV model here.  The
temperature evolution, on the other hand, is governed by
$\hat{z}\times \nabla T \cdot \nabla B$ nonlinearity that is identical
to a diamagnetic nonlinear term in HMW model. The role of this
nonlinearity has traditionally been identified as a source of
suppressing the intensity of nonlinear flows in drift wave turbulence.
Nevertheless, the presence of the linear inhomogenous background in
both equations can modify the nonlinear mode coupling interactions
in a subtle manner. Our objective is to understand the latter in the
context of nonlinear interactions mediated by the inhomogeneous $B$
and $T$ fields in MEDV modes.

We find from our small amplitude nonlinear evolution case (see Fig 1)
that the nonlinear interactions in the inhomogeneous MEDV modes are
typically led by the Reynolds-stress-like nonlinearity,
i.e. $\hat{z}\times \nabla B \cdot \nabla \nabla^2 B$, in the magnetic
field equation. Consequently, the mode structures in the saturated
turbulent state are dominated by the zonal-like flows, as typically
observed in the Hasegawa-Mima-Wakatani (HMW) model
\cite{r13,r14,r15,r16}. This is demonstrated in Fig. (1) where large
scale structures (zonal flow-like, $k_y \approx 0$) along the background
magnetic and temperature field gradients are developed in $B, T,
\nabla^2 B$ and $B-\nabla^2 B$ fluctuations. Thus, small amplitude
nonlinear evolution, as shown in Fig. (1), is consistent with the HMW model
of the inhomogeneous drift-wave turbulence. Our simulations described in
Fig. (1) can be contrasted with our previous work \cite{Shaikh08} on
the small amplitude case that was studied in the absence of a
background magnetic field gradient. In Ref \cite{Shaikh08}, we have
shown that mode coupling interactions during the nonlinear stage of
evolution leads to the formation of streamer-like structures in the
magnetic field fluctuations associated with $k_y\approx 0, k_x \ne
0$. These structures were similar to the zonal flows but contained a
rapid $k_x$ variations, thus the corresponding frequency is relatively
large.  The temperature fluctuations in Fig (1) of \cite{Shaikh08}, on
the other hand, depict an admixture of isotropically localized
turbulent eddies and a few stretched along the direction of the
background inhomogeneity. Clearly, the presence of a background
magnetic field gradient reduces the $k_x$ variation in the steady-state
flow. This leads to a considerable modification in the final
structures, which now appear to look like zonal-flows (and not the
streamers that were observed in the absence of a background magnetic
field). Interestingly, the large amplitude case in our simulations
unravels a completely different scenario where nonlinear mode coupling
interactions in MEDV modes are observed to suppress the steady-state
flow. This is shown in Fig. (2). The suppression of the zonal
flow-like structures can be understood in the context of the
diamagnetic-like nonlinearity that seems to modify the nonlinear mode
coupling interactions in the large amplitude evolution. This
nonlinearity, corresponding to a $\{T, B\}$ term in the temperature
equation, becomes gradually strong enough to nullify the emergence of
the large scale zonal-flow-like structures. Hence, the steady state
structures in Fig. (2) appear to possess more small scale fully
developed turbulent fluctuations. We confirm that this state is not
entirely isotropic and the background gradients nonlinearly maintain
the anisotropic cascades in inhomogeneous MEDV modes. We elucidate
this point in the following section.

\begin{figure}[t]
\begin{center}
\includegraphics[width=10.cm,height=8.cm]{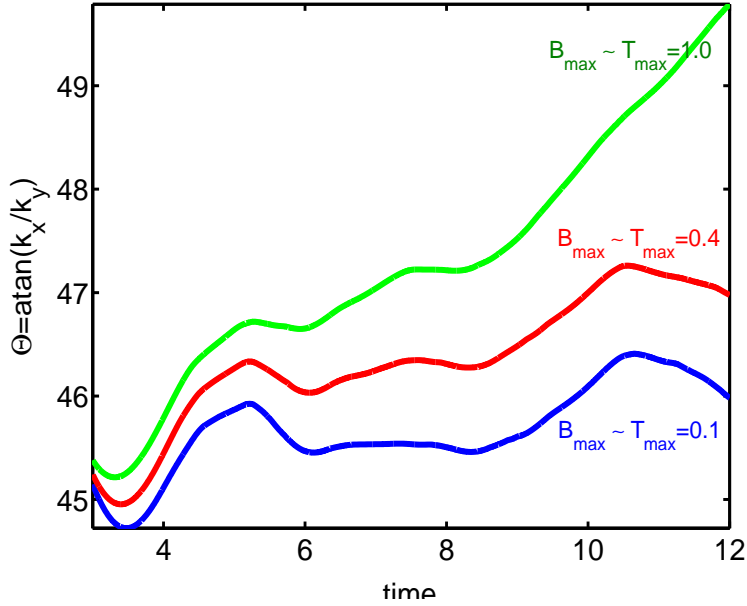}
\end{center}
\caption{\label{fig3} (Color online) Evolution of anisotropic mode structures as
  described by $k_x$ and $k_y$ mode averaged over the entire turbulent
  spectrum in inhomogeneous MEDV turbulence. Initially, $k_x =
  k_y$. Progressive development of anisotropy $k_x > k_y$ is ascribed
  to the presence of background gradients in magnetic and temperature
  fields for which the anisotropy angle $\Theta=\tan^{-1}(k_x/k_y)$
  deviates continually from $45^\circ$.}
\end{figure}

\section{Anisotropic MEDV cascades}

We quantify the degree of anisotropy mediated by the presence of large
scale gradients in the magnetic and temperature fields in the nonlinear 2D
inhomogeneous MEDV turbulence. In 2D turbulence, the anisotropy in the
$k_x-k_y$ plane is associated with the preferential transfer of
spectral energy that empowers either of the $k_x$ and $k_y$ modes. The
anisotropy in the initial isotropic turbulent spectrum is triggered
essentially by the background anisotropic gradients that nonlinearly
migrate the spectral energy in a particular direction. To measure the
degree of anisotropic cascades, we employ the following diagnostics to
monitor the evolution of $k_x$ mode in time. The $k_x$ mode is
determined by averaging over the entire turbulent spectrum that is
weighted by $k_x$.
\[ k_x(t) = \sqrt{\frac{\sum_k |k_x Q(k,t)|^2}{\sum_k |Q(k,t)|^2}} \]
Here $Q$ represents any of $B$, $T$, $\nabla^2 B$ and $B-\nabla^2 B$.
Similarly, the evolution of $k_y$ mode is determined by the following
relation.
\[ k_y(t) = \sqrt{\frac{\sum_k |k_y Q(k,t)|^2}{\sum_k |Q(k,t)|^2}} \]
We can define an angle of anisotropy such that
$\Theta=\tan^{-1}(k_x/k_y)$.  It is clear from these expressions that
the $k_x$ and $k_y$ modes exhibit isotropy when $k_x \simeq k_y$ for
which $\Theta \simeq 45^\circ$.  Any deviation from this equality
leads to a spectral anisotropy. We follow the evolution of $k_x$ and
$k_y$ modes in our simulations for increasing amplitude of the initial
fluctuations.  Our simulation results describing the evolution of
$k_x$ and $k_y$ modes are shown in Fig. 3. It is evident from Fig. 3
that the initial isotropic modes $k_x \simeq k_y$ gradually evolve
towards an highly anisotropic state in that spectral transfer
preferentially occurs in the $k_x$ mode, while the same is suppressed
in $k_y$ mode. Consequently, the spectral transfer in $k_x$ mode
dominates the evolution and the mode structures show elongated
structures along the $y$-direction. With the increasing amplitude of
the initial fluctuations, there exists increasing degree of anisotropy
as illustrated in Fig 3. The larger the amplitude is, stronger are the
nonlinear interactions. Correspondingly, there exists increasing
degree of disparity in the $k_x$ and $k_y$ modes owing primarily to
the increasing depletion in $k_y$ modes as shown in Fig 3.

\begin{figure}[t]
\begin{center}
\includegraphics[width=10.cm,height=8.cm]{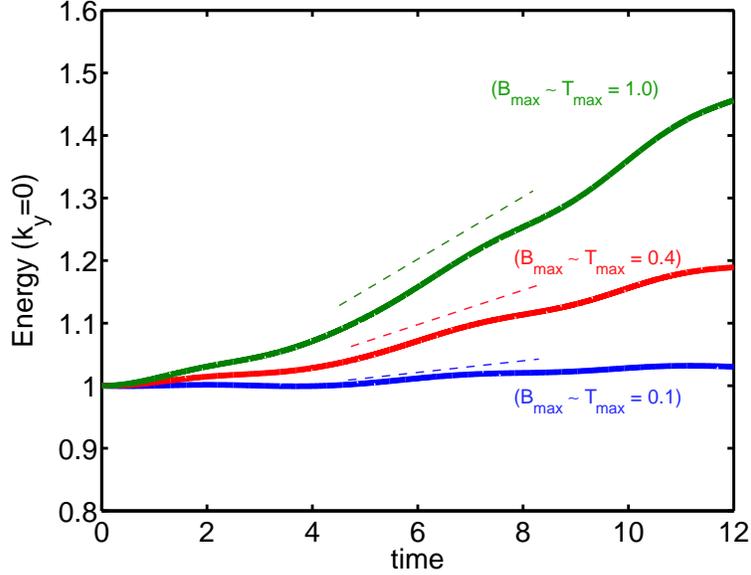}
\end{center}
\caption{\label{fig4} (Color online) Energy associated with the anisotropic flows
  that corresponds essentially to the nonlinear $k_y=0$ modes. Figure
  depicts the self-consistently generated nonlinear flows lead to
  turbulent anisotropy. The growth rate of the generation of
  anisotropic flow is directly proportional to the amplitude of the
  initial fluctuations.}
\end{figure}

The increasing angle of anisotropy (in Fig 3), with the increasing
magnitude of the initial fluctuations, is ascribed to the generation
of highly asymmetric flows in our simulations. We exemplify this point
by means of Fig 4 that describes evolution of the energy associated
with the anisotropic flows led predominantly by the $k_y=0$
mode. Notably, this mode is generated explicitly by the nonlinear
interactions. It is clear from this figure that growth rate of the
generation of the anisotropic (dominated by the $k_y=0$ mode) flow is
directly proportional to the magnitude of the initial fluctuations.
Thus the growth rate is higher when the amplitude of the magnetic and
temperature field is large, i.e. $B_{max} \simeq T_{max} = 1.0$.  The
smaller initial amplitude of the fluctuations correspond to relatively
weak nonlinear interactions for which the generation of the $k_y=0$
mode is insignificant.  The nonlinear interactions do not introduce
anisotropy and hence lead to nearly isotropic turbulent cascades. This
is observed clearly in Fig 4 (see the curve corresponding to $B_{max}
\simeq T_{max} = 0.1$) which is consistent with the corresponding
curve in Fig 3. In Fig 4, each curve is normalized with it's own
initial value to rescale all the curves on a single plot. This further
enables us to make a vis a vis comparison between the three different
cases shown in Fig 4.  Also evident from the figure 4 is the large
amplitude fluctuations that lead to the higher growth rate of the
anisotropic flows (see dashed lines whose slope increases with the
increasing amplitude of $B_{max}$ and $T_{max}$). The increasing
slope, i.e. growth, associated with the anisotropic flows in our
simulations is further consistent with the increasing angle of
anisotropy that is observed in Fig 3.

\begin{figure}[t]
\begin{center}
\includegraphics[width=10.cm,height=8.cm]{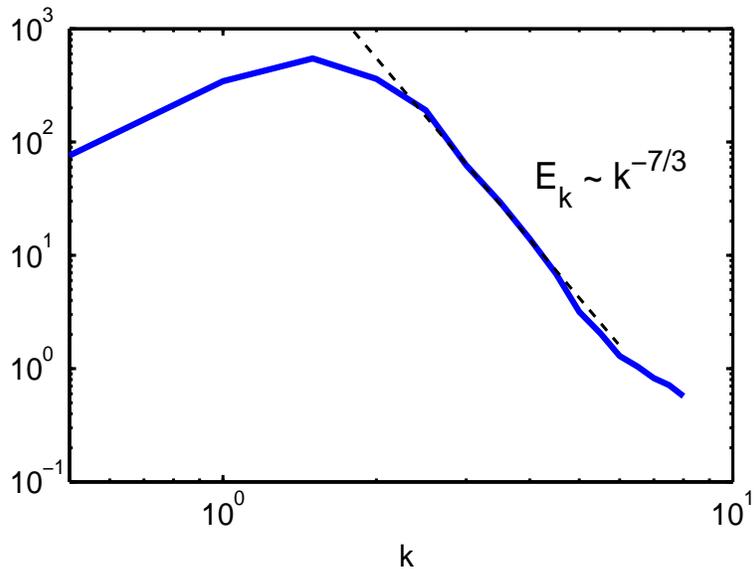}
\end{center}
\caption{\label{fig5} (Color online) Anisotropic inhomogeneous MEDV turbulent
  fluctuations exhibit a $k^{-7/3}$ like spectrum. The spectrum
corresponds to an intermittent state where turbulence coexists with
anisotropic structures.}
\end{figure}

It is noteworthy from our simulations that the stronger nonlinear
interactions pile up an increasing amount of turbulent energy in the
$k_y=0$ mode. Consequently, the turbulent correlation length scales
tend to decrease across the flows. Hence the large amplitude
simulations show decorrelated flow structures in Fig 2.  This
physically means that the nonlinear interactions led by the
polarization and diamagnetic like terms in the presence of background
gradients quench the flow (that was observed along the $x$-direction
in Fig. 1) to introduce reduced turbulent decorrelated structures.

While there exists a disparity in the spectral transfer of energy
corresponding to the $k_x$ and $k_y$ modes, the 2D volume averaged
turbulent spectrum follows a $k^{-7/3}$ power law, as shown in
Fig. (5).  This spectrum is steeper than that of the HMW turbulence
\cite{r13,r14,r15,r16}.  The steepness of the observed spectrum can be
ascribed to the coexistence of partially anisotropic flows and
turbulent fluctuations in the steady state MEDV
mode turbulence.

\section{Summary and conclusions}

In summary, we have investigated the properties of the MEDV mode
turbulence nonuniform magnetoplasma containing gradients in the
electron temperature, the electron number density, and the external
magnetic field.  We find that the influence of the magnetic field
gradient is to suppress the streamer--like structures observed in
simulations without the magnetic field gradient \cite{Shaikh08}. In
addition, the steep spectra changes to spectra with a 7/3 power law in
the presence of the magnetic field gradient. We also discussed the
conditions for an instability through a first order baroclinic effect,
which would lead to the generation of magnetic fields under conditions
where the background density and temperature gradients are in the same
direction.  It is noteworthy that the anisotropic terms in our
simulations become important when the linear terms, corresponding to
the gradients in magnetic and temperature fields, compete with the
nonlinear terms.  In such case, efficient migration of energy takes
place between the gradients and turbulent modes that primarily lead to
the nonlinear anisotropic flows.  Our study could have relevance for
laser produced plasmas in the laboratory, spontaneously generated
magnetic fields, and effects associated with self-generated magnetic
fields such as transport of energy along surfaces and flux limitation
of electrons, have been observed for many years.

{\bf Acknowledgments} This research was partially supported by the
Swedish Research Council (VR) and by the Deutsche
Forschungsgemeinschaft through the Forschergruppe FOR 1048.  Dastgeer
Shaikh acknowledges the support of NASA(NNG-05GH38) and NSF
(ATM-0317509) grants.

\end{document}